\pdfoutput=1
\RequirePackage{ifpdf}
\ifpdf 
\documentclass[pdftex]{sigma}
\else
\documentclass{sigma}
\fi

\usepackage{bm}
\usepackage{bbm}
\usepackage{epsf}
\usepackage{nicefrac}

\newcommand{\dd}{\mathrm{d}}
\newcommand{\ee}{\mathrm{e}}
\newcommand{\ii}{\mathrm{i}}

\begin{document}

\allowdisplaybreaks

\renewcommand{\thefootnote}{$\star$}

\renewcommand{\PaperNumber}{005}

\FirstPageHeading

\ShortArticleName{Nonanalytic Expansions and Anharmonic Oscillators}

\ArticleName{Generalized Nonanalytic Expansions,\\ $\boldsymbol{\mathcal{PT}}$-Symmetry  and
Large-Order Formulas\\ for Odd Anharmonic Oscillators\footnote{This paper is a contribution to the Proceedings of the VIIth Workshop ``Quantum Physics with Non-Hermitian Operators''
     (June 29 -- July 11, 2008, Benasque, Spain). The full collection
is available at
\href{http://www.emis.de/journals/SIGMA/PHHQP2008.html}{http://www.emis.de/journals/SIGMA/PHHQP2008.html}}}

\Author{Ulrich D. JENTSCHURA~$^{\dag^1}$,
Andrey SURZHYKOV~$^{\dag^2}$ and Jean ZINN-JUSTIN~$^{\dag^3}$}

\AuthorNameForHeading{U.D. Jentschura, A. Surzhykov and J. Zinn-Justin}

\Address{$^{\dag^1}$~Department of Physics, Missouri University of Science
and Technology,\\
\hphantom{$^{\dag^1}$}~Rolla MO65409-0640, USA}
\EmailDD{\href{mailto:ulj@mst.edu}{ulj@mst.edu}}

\Address{$^{\dag^2}$~Physikalisches Institut der Universit\"{a}t,
Philosophenweg 12, 69120 Heidelberg, Germany}

\Address{$^{\dag^3}$~CEA, IRFU and Institut de Physique
Th\'{e}orique, Centre de Saclay,\\
\hphantom{$^{\dag^3}$}~F-91191 Gif-Sur-Yvette, France}

\ArticleDates{Received October 30, 2008, in f\/inal form January 07,
2009; Published online January 13, 2009}

\Abstract{The concept of a generalized nonanalytic expansion
which involves nonanalytic combinations of exponentials,
logarithms and powers of a coupling is introduced and its use
illustrated in various areas of physics.
Dispersion relations for the resonance energies of odd anharmonic
oscillators are discussed, and higher-order formulas are presented
for cubic and quartic potentials.}


\Classification{81Q15; 81T15}

\section{Introduction and motivation}

In many cases, a simple power series, which may be the result
of a Taylor expansion,  is not enough in order to
describe a physical phenomenon.
Furthermore, even if a power series expansion
(e.g., of an energy level in terms of some coupling parameter)
is possible, then it may not be
convergent~\cite{BeWu1969,BeWu1971,BeWu1973,LGZJ1977,LGZJ1980}.
Physics is more complicated,
and generalizations of the concept of a~simple Taylor
series are called for.

Let us start with a simple example, an electron
bound to a nucleus. It is described to a
good accuracy by the Dirac equation involving the
Dirac--Coulomb (DC) Hamiltonian,
\begin{equation*}
H_{\mathrm{DC}} \psi = E_{\mathrm{DC}}   \psi  ,
\qquad
H_{\mathrm{DC}} = \vec{\alpha}\cdot\vec{p} + \beta m - \frac{Z\alpha}{r}  .
\end{equation*}
Here, natural units ($\hbar = c = \epsilon_0 = 1$)
are employed, and the
familiar Dirac matrices are denoted by the symbols
$\vec{\alpha}$ and $\beta$.
The energy of an $nS_{1/2}$ state
(we use the usual spectroscopic notation for the quantum numbers)
reads, when expanded up to
sixth order in the parameter $Z\alpha$,
\begin{gather*}
E_{\mathrm{DC}} =  m - \frac{(Z\alpha)^2   m}{2   n^2}
- \frac{(Z\alpha)^4   m}{n^3}
\left( \frac12 - \frac{3}{8  n} \right)
\nonumber\\
\phantom{E_{\mathrm{DC}} =}{}  + \frac{(Z\alpha)^6   m}{n^3}
\left( -\frac18 - \frac{3}{8 n} + \frac{3}{4  n^2} -
\frac{5}{16   n^3} \right)
+ \mathcal{O}(Z\alpha)^6  .
\end{gather*}
This is a power expansion in the parameter $Z\alpha$, where
$Z$ is the nuclear charge number and $\alpha$ is the f\/ine-structure
constant, and for $Z\alpha < 1$, it converges to the well-known
exact Dirac--Coulomb eigenvalue~\cite{ItZu1980}.

On the other hand, let us suppose, hypothetically,
that the electron were to carry
no spin. Then, the equation would change to the
bound-state equation for a Klein--Gordon particle,
\begin{equation*}
H_{\mathrm{KG}} \psi = E_{\mathrm{KG}}  \psi ,
\qquad
H_{\mathrm{KG}} = \sqrt{\vec{p}\,{}^{2} + m^2} - \frac{Z\alpha}{r}  .
\end{equation*}
In the expansion of an $S$-state energy levels in terms of
$Z\alpha$, an irregularity develops for
spinless particles, namely, a $(Z\alpha)^5$ term, and the
$(Z\alpha)^6$ term carries a logarithm
(see~\cite{Pa1997} for a detailed derivation):
\begin{gather*}
E_{\mathrm{KG}} =  m - \frac{(Z\alpha)^2 m}{2 n^2}
- \frac{(Z\alpha)^4 m}{n^3}
\left( 1 - \frac{3}{8 n} \right)
+ \frac{8 \, (Z\alpha)^5 m}{3 \, \pi \, n^3}
+ \frac{(Z\alpha)^6 m}{n^3} \Bigg[ \ln(Z\alpha)
+ \frac{7}{\pi^2} \zeta(3)
\nonumber\\
\phantom{E_{\mathrm{KG}} =}{} - \frac{2}{\pi^2}
+ \Psi(n) + \gamma_{\mathrm{E}} - \ln(n) - \frac{1}{n} +
\frac{5}{3 \, n^2}  - \frac{5}{16\, n^3} - \frac{29}{12} \Bigg]
+ \mathcal{O}(Z\alpha)^6 .
\end{gather*}
The expansion is nonanalytic (we denote by $\Psi$ the
logarithmic derivative of the Gamma function, and $\gamma_{\mathrm{E}}$
is Euler's constant). The occurrence of nonanalytic terms
has been key not only to~gene\-ral bound-state
calculations, but in particular also to Lamb shift calculations,
which entail nonanalytic expansions
in the electron-nucleus coupling strength
$Z\alpha$ in addition to power series
in the quantum electrodynamic (QED) coupling $\alpha$.
A few anecdotes and curious stories are
connected with the evaluation of higher-order
logarithmic corrections to the Lamb
shift~\cite{ErYe1965a,ErYe1965b,Ka1996,Pa2001}.
The famous and well-known Bethe logarithm,
by the way, is the nonlogarithmic (in~$Z\alpha$)
part of the energy shift in the order $\alpha   (Z\alpha)^4$,
and it is a subleading term following the
leading-order ef\/fect which is of the functional
form $\alpha   (Z\alpha)^4   \ln(Z\alpha)$.

It does not take the additional complex structure of a
Lamb shift calculation to necessitate the introduction of logarithms,
as a simple model example based on an integral
demonstrates~\cite{JePa2002},
\begin{gather*}
\int_0^1 \sqrt{\frac{\omega^2 + \beta^2}{1 - \omega^2}}
{\mathrm{d}}\omega
\; \mathop{=}^{\beta \to 0} \;  1 + \beta^2
\left\{ \frac{1}{2}
\ln\left(\frac{4}{\beta}\right) + \frac{1}{4} \right\}
+ \beta^4   \left\{ -\frac{1}{16}
\ln\left(\frac{4}{\beta}\right) + \frac{3}{64} \right\}
\nonumber\\
\hphantom{\int_0^1 \sqrt{\frac{\omega^2 + \beta^2}{1 - \omega^2}}
{\mathrm{d}}\omega
\; \mathop{=}^{\beta \to 0} \;}{}  + \beta^6   \left\{ \frac{3}{128}
\ln\left(\frac{4}{\beta}\right) - \frac{3}{128} \right\}
+ {\mathcal{O}}(\beta^8   \ln \beta) .
\end{gather*}
Another typical functional form in the description
of nature, characteristic of tunneling pheno\-mena,
is an exponential factor. Let us consider, following
Oppenheimer~\cite{Op1928}, a hydrogen atom in an external electric
f\/ield (with f\/ield strength $|\vec E|$).  The nonperturbative
decay width due to tunneling is proportional to
\begin{equation*}
\exp\left[- \frac{2 (Z\alpha)^3 m}{ 3 |e \vec E|} \right] ,
\end{equation*}
where $| e \vec E |$ is the modulus of the electron's electric charge
multiplied by the static electric f\/ield strength.

We have by now encountered three functional forms
which are typically necessary in order to describe expansions
of physical quantities: these are
simple powers, which are due to higher-order perturbations
in some coupling parameter, logarithms due to some cutof\/f,
and nonanalytic exponentials. The
question may be asked as to whether phenomena exist
whose description requires the use of all three mentioned functional
forms within a single, generalized nonanalytic expansion?

The answer is af\/f\/irmative, and indeed, for the description of energy levels of
the double-well potential, it is known that we have to invoke a triple
expansion in the quantities $\exp(-A/g)$, $\ln(g)$ and powers of $g$ in order
to describe higher-order ef\/fects~\cite{ZJJe2004i,ZJJe2004ii}
(here, $g$ is a coupling parameter which is roughly proportional
to the inverse distance of the two minima of the double-well
potential). Other
potentials, whose ground-state energy has a vanishing perturbative expansion to
all orders (e.g., the Fokker--Planck potential), also can be described using
generalized expansions~\cite{JeZJ2004plb}.  The double-well and the
Fokker--Planck Hamiltonians have stable, real eigenva\-lues when
acting on the Hilbert space of square-integrable wave functions
(no complex resonance eigenvalues). An interesting class of
recently studied potentials is
$\mathcal{PT}$-symmetric~\cite{BeBo1998,BeDu1999,BeBoMe1999,BeBrJo2002}.
Odd anharmonic oscillators for imaginary coupling fall into this
class, but the double-well and the
Fokker--Planck Hamiltonians do not.
The purpose of this contribution is to assert that the concept
of $\mathcal{PT}$-symmetry is helpful as an auxiliary device
in the study of odd anharmonic oscillators.

In contrast to our recent investigation~\cite{JeSuZJ2009prl},
we here focus on a few subtle issues associated with the
formulation of the dispersion relation for odd anharmonic
oscillators (Section~\ref{toward}), before giving a few new results
for the cubic and quartic anharmonic oscillators in
Section~\ref{some_results}. In~\cite{JeSuZJ2009prl},
by contrast, we focus on the sextic and
septic oscillators. Conclusions are reserved for Section~\ref{conclu}.

\section{Toward anharmonic oscillators}
\label{toward}

Let us brief\/ly recall why it is nontrivial to write
dispersion relations for the energy levels of odd anharmonic
oscillators. We consider as an example an odd perturbation of the form
$\gamma x^3$, with a~coupling parameter $\gamma$,
and we emphasize the dif\/ferences to even anharmonic oscillators.
Let us therefore investigate, as a function of the
coupling parameter $\gamma$, the quartic and cubic potentials
$u(\gamma, x) = \tfrac12 x^2 + \gamma   x^4$ and
$v(\gamma, x) = \frac12 x^2 + \gamma   x^3$.

\begin{figure}[t]
\begin{center}
\begin{minipage}[b]{0.35\linewidth}
\begin{center}
\includegraphics[width=1.0\linewidth,angle=0, clip=]{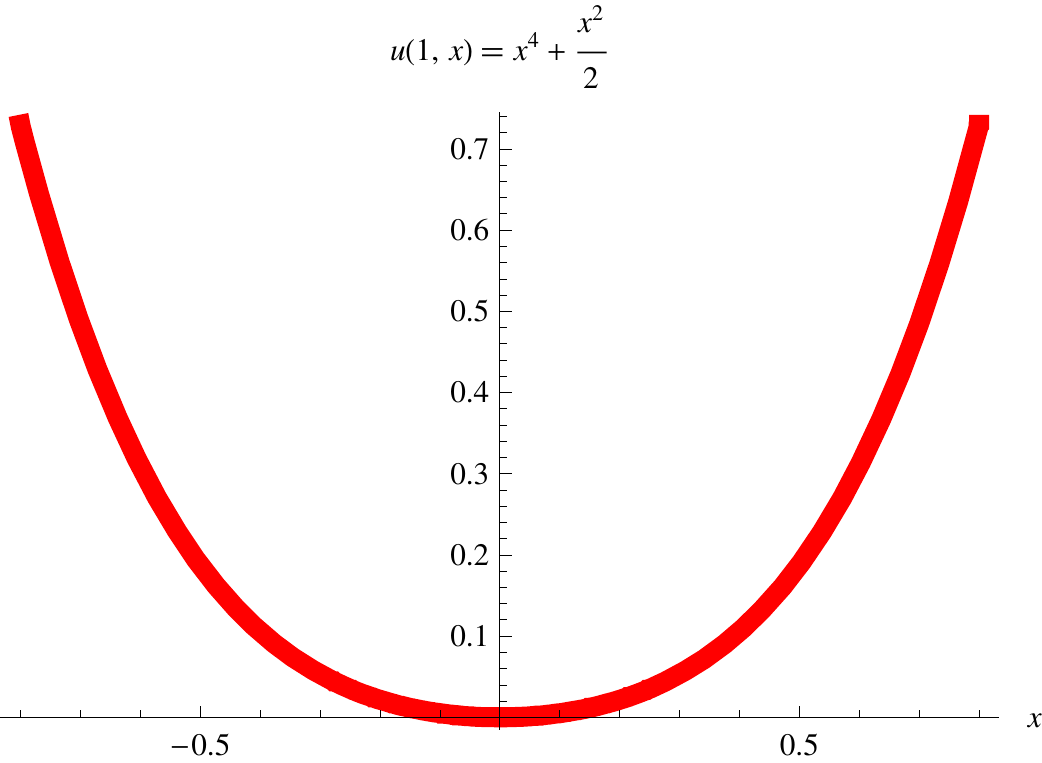} (a)
\end{center}
\end{minipage}
\hspace*{1cm}
\begin{minipage}[b]{0.35\linewidth}
\begin{center}
\includegraphics[width=1.0\linewidth,angle=0, clip=]{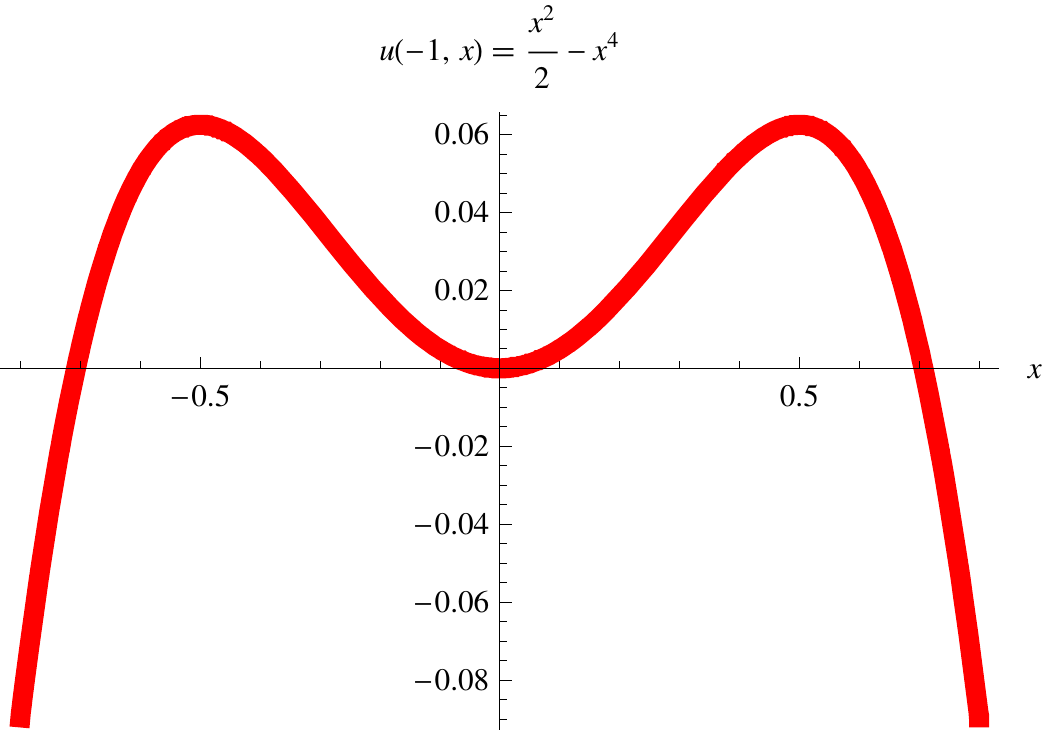} (b)
\end{center}
\end{minipage}
\end{center}

\begin{center}
\begin{minipage}[b]{0.35\linewidth}
\begin{center}
\includegraphics[width=1.0\linewidth,angle=0, clip=]{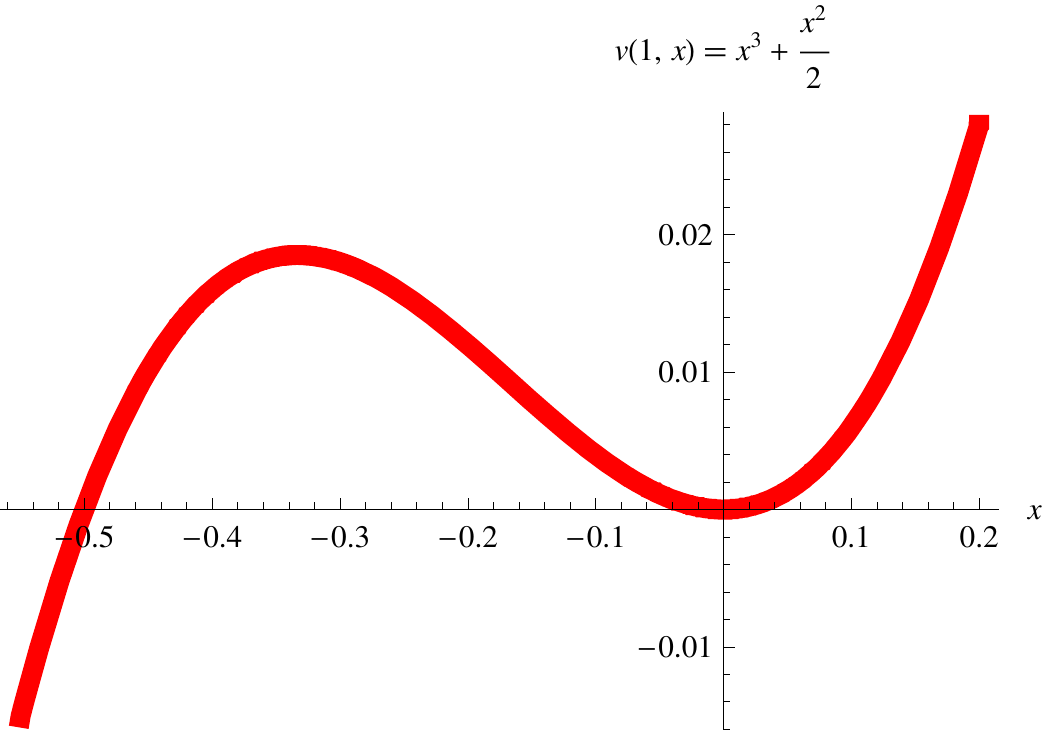} (c)
\end{center}
\end{minipage}
\hspace*{1cm}
\begin{minipage}[b]{0.35\linewidth}
\begin{center}
\includegraphics[width=1.0\linewidth,angle=0, clip=]{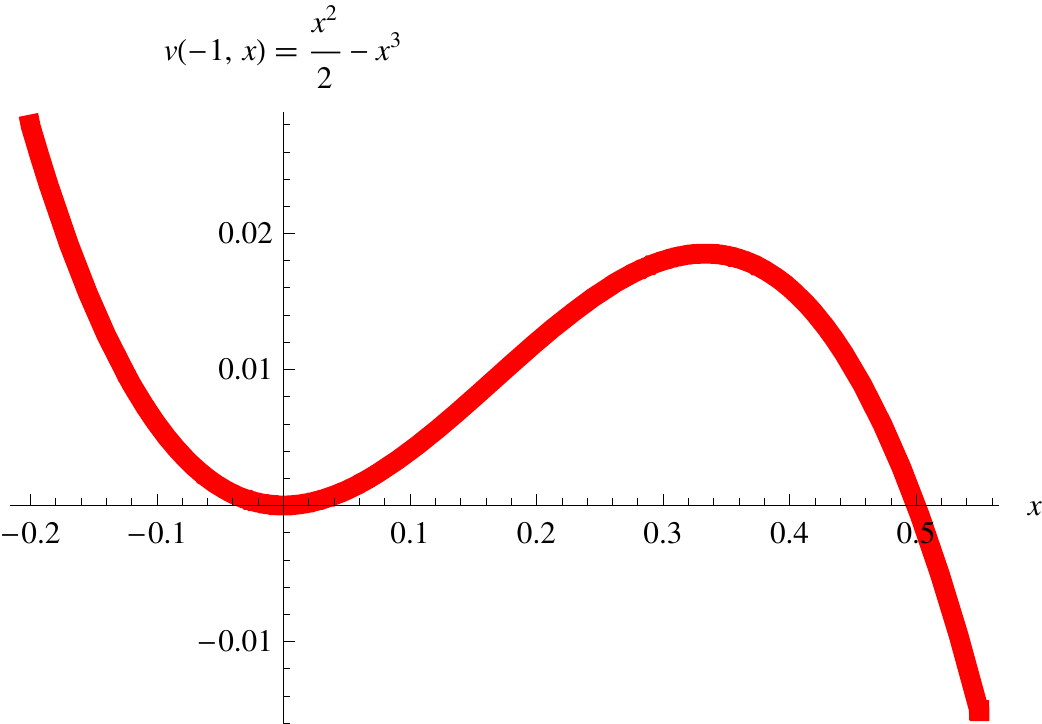} (d)
\end{center}
\end{minipage}
\end{center}

\caption{Plot of the potentials
$u(\gamma, x) = \tfrac12 x^2 + \gamma   x^4$ and
$v(\gamma, x) = \tfrac12 x^2 + \gamma   x^3$
for positive and negative coupling $\gamma = \pm 1$.}\label{fig1}
\end{figure}

For the quartic potential and positive coupling
$\gamma > 0$, the spectrum of the Hamiltonian
$-\frac12   \partial_x^2 + u(\gamma > 0, x)$, endowed with
$L^2$ boundary conditions,
consists of discrete energy levels which represent physically
stable states with zero decay width.
For $\gamma < 0$, the potential $u(\gamma, x)$
has a double-hump structure,
and the particle can escape either to the left or to the right
of the ``middle valley'' by tunneling
(see Figs.~\ref{fig1}(a) and~\ref{fig1}(b)).
So, when we change the sign of the coupling parameter,
then ``the physics of the potential changes drastically.''
We can then use the fact that, as a function of $\gamma$,
the energy eigenvalues of the quartic oscillator
have a branch cut along the negative real
axis~\cite{BeWu1969,BeWu1971,BeWu1973} and write a
dispersion relation. It has been stressed
in~\cite{ZJ1996,ZJ2003} that the discontinuity of the energy
levels is given exactly by the
instanton conf\/iguration, and this fact has been widely used
in the literature in the analysis of related problems in
quantum physics and f\/ield theory.

(Actually, when acting on $L^2$, the negative-coupling
quartic potential still possesses a real spectrum with
discrete eigenvalues, but the analysis is highly
nontrivial~\cite{FePe1995}.
Indeed, the natural eigenenergies that are obtained
from the real energies for positive coupling by analytic
continuation as the complex argument of coupling parameter
and of the boundary conditions, are just the
complex resonance energies for which the dispersion relation holds.)

Now let us investigate  the odd potential
$v(\gamma, x) = x^2/2 + \gamma   x^3$.
When $\gamma$ here changes sign, the physics of the
potential does not change (see Figs.~\ref{fig1}(c)
and~\ref{fig1}(d)): still, the particle can escape
the ``middle valley'' by tunneling.
Resonances occur. The question is whether we now have two
branch cuts as a function of $\gamma$, one along the
positive-$\gamma$ axis and one for negative $\gamma$.
Should we attempt to formulate
a dispersion with integration along
$\gamma \in (-\infty, 0)$ and
$\gamma \in (0, \infty)$? The answer is no.
Rather, we should redef\/ine the coupling in such a way that
the $\mathcal{PT}$-symmetry of the potential is
used ef\/fectively. This means that the spectrum is real
for purely imaginary coupling $\gamma = \ii \, \beta$
with real $\beta$, and it is
invariant under the transformation $\gamma = \ii \,\beta \to
\gamma = -\ii \,\beta $.
In some sense, the case of the cubic potential
for purely imaginary coupling is equivalent to the quartic potential
for positive coupling parameter,
and the case of the cubic potential
for positive coupling is equivalent to the quartic potential
for negative coupling parameter.
The key thus is to formulate the energy levels of the
cubic as a function of $g = \gamma^2$, not $\gamma$
itself~\cite{BeDu1999}.

\section{Some results}
\label{some_results}

\begin{figure}[t]
\begin{minipage}[b]{0.49\linewidth}
\begin{center}
\includegraphics[width=1.0\linewidth,angle=0, clip=]{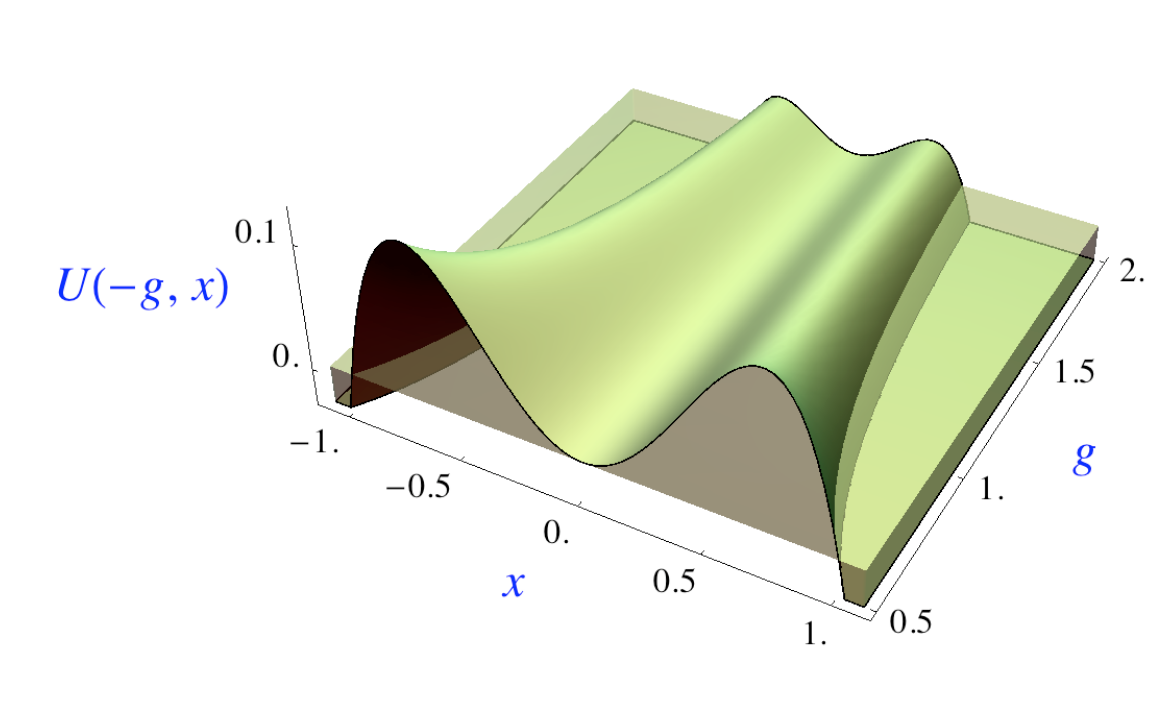} \\ (a)
\end{center}
\end{minipage}
\begin{minipage}[b]{0.49\linewidth}
\begin{center}
\includegraphics[width=1.0\linewidth,angle=0, clip=]{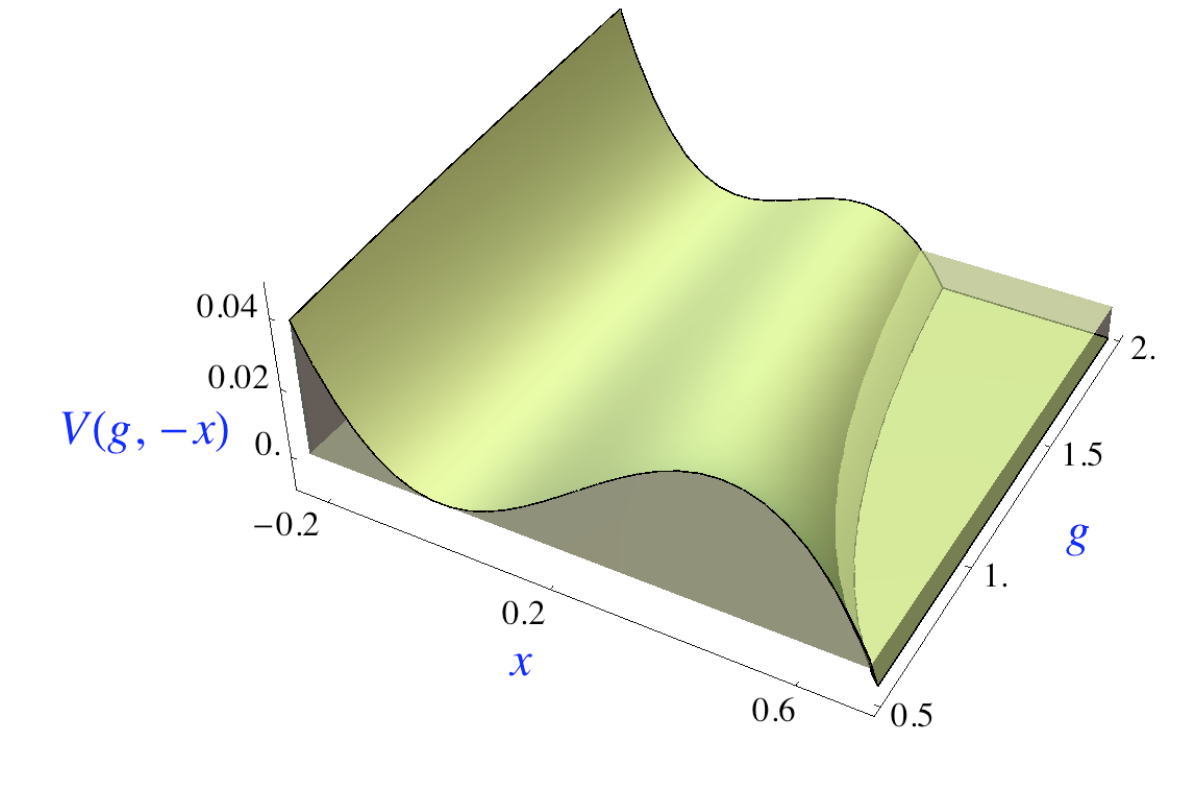} \\ (b)
\end{center}
\end{minipage}

\caption{(a) The quartic potential
$U(g, x) = \frac12  x^2 + g  x^4$ is
plotted in the parameter range $g \in (-2.0, -0.5)$.
As a function
of the coupling $g$, the distance that the quantal
particle has to tunnel before it escapes from the
middle valley decreases as the coupling parameter
increases. The decay width of the ground state
is proportional to $\exp[1/(3g)]$ (for $g < 0$) and increases
with the coupling. Higher-order corrections to this
well-known result are indicated here (see equation~(\ref{higher4})).
(b) The cubic potential
$V(g, x) = \frac12   x^2 + \sqrt{g}   x^3$ is
plotted in the parameter range $g \in (0.5, 2.0)$.
The decay width of the ground state
is proportional to $\exp[-2/(15 g)]$ and increases
with the coupling.}\label{fig2}
\end{figure}

We here summarize a few results obtained recently~\cite{JeSuZJ2009prl}
regarding the higher-order corrections for the energy levels of
general even-order and odd-order anharmonic oscillators,
using the quartic and cubic potentials as examples.
Let us thus consider the two Hamiltonians,
\begin{alignat*}{3}
& H(g) =  - \frac12  \frac{\partial^2}{\partial x^2} +
U(g, x)  ,
\qquad &&
U(g, x) = \frac12   x^2 + g   x^4  ,&
\\
& h(g) =  - \frac12  \frac{\partial^2}{\partial x^2} +
V(g, x)  ,
\qquad
&&
V(g, x) = \frac12   x^2 + \sqrt{g}   x^3  ,
\end{alignat*}
in the unstable region, i.e.~for $g < 0$ in the
quartic case, and for $g > 0$ in the cubic
case. We assume both Hamiltonians to be
endowed with boundary conditions
for the resonance energies (which leads to a
nonvanishing negative imaginary part for the
resonance energy eigenvalues).
Specif\/ically, we denote the resonance eigenenergies by
$E_n(g)$ for the quartic and $\epsilon_n(g)$
for the cubic, respectively.
The quartic potential
is plotted in the range $g \in (-2, -\tfrac12)$ in
Fig.~\ref{fig2}(a), and the cubic potential
is plotted in the range $g \in (\tfrac12, 2)$
in Fig.~\ref{fig2}(b).

\begin{figure}[t]
\begin{minipage}[b]{0.49\linewidth}
\begin{center}
\includegraphics[width=1.0\linewidth,angle=0,clip=]{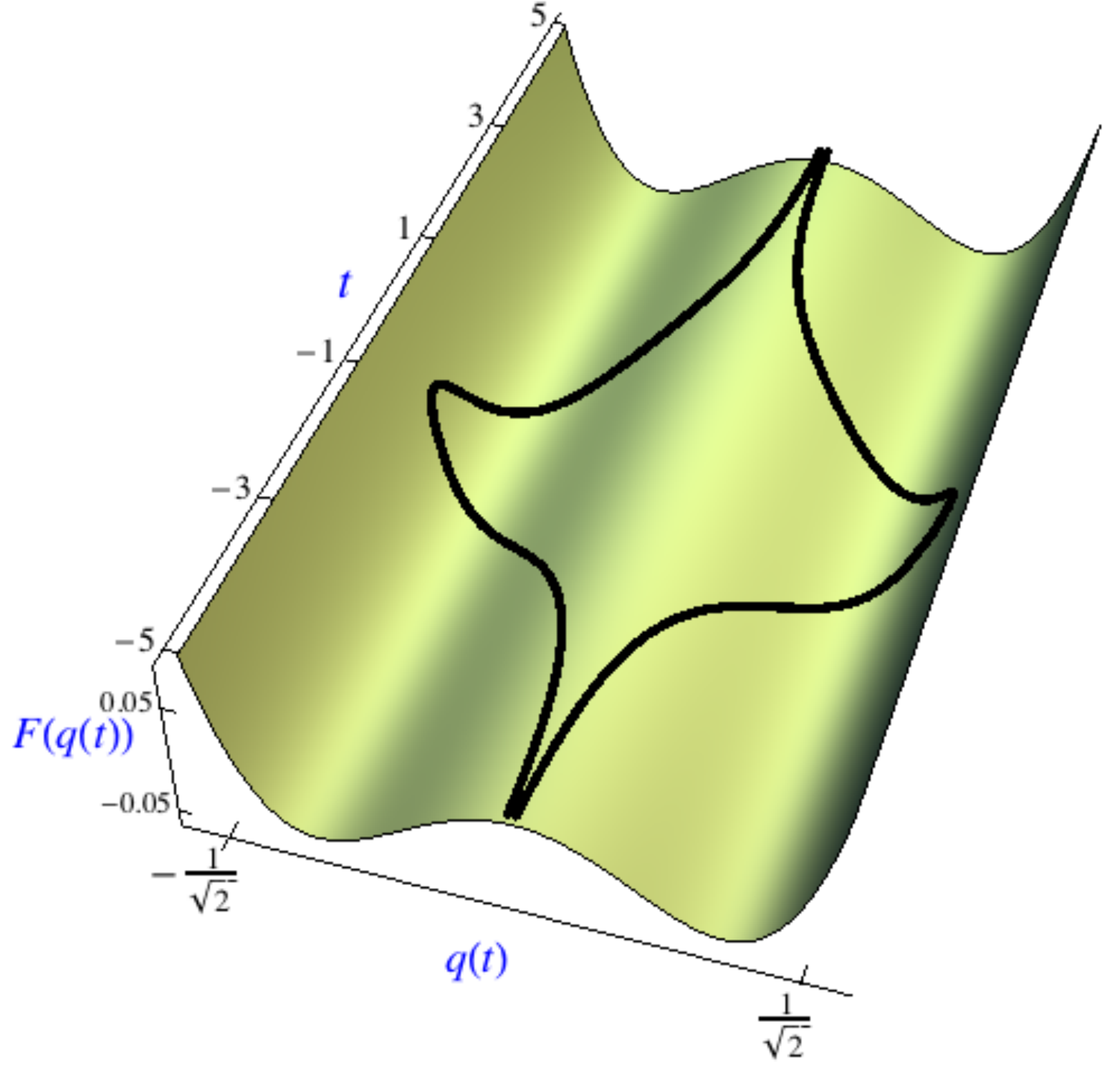} \\ (a)
\end{center}
\end{minipage}
\begin{minipage}[b]{0.46\linewidth}
\begin{center}
\includegraphics[width=1.0\linewidth,angle=0,clip=]{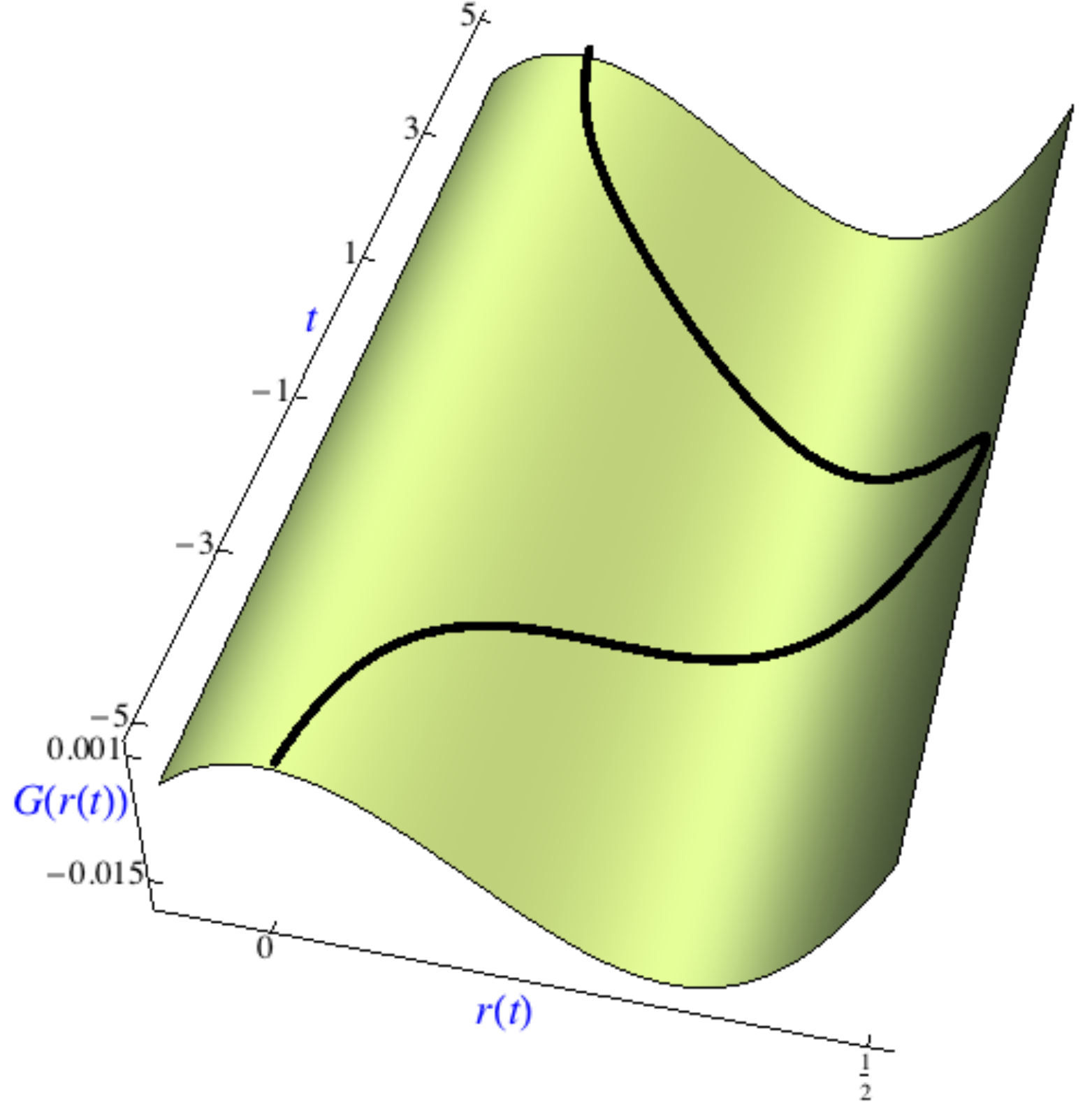} \\ (b)
\end{center}
\end{minipage}
\caption{(a) Quartic instanton. (b) Cubic instanton.
See also equations~(\ref{inst4}) and (\ref{inst3}).}\label{fig3}
\end{figure}

Let us now investigate the instanton actions
(see also Fig.~\ref{fig3}).
We write the classical Euclidean actions for the
quartic and cubic, respectively, as
\begin{equation*}
S[x] = \int \dd t   \left(
\tfrac12   \dot x^2 +
\tfrac12   x^2 + g\, x^4 \right)  , \qquad
s[y] = \int \dd t   \left(
\tfrac12   \dot y^2 +
\tfrac12   y^2 + \sqrt{g}   y^3 \right)  ,
\end{equation*}
and perform the following scale transformation
$x \equiv x(t) = (-g)^{-1/2} \, q(t)$ and
$y \equiv y(t) = -g^{-1/2} \, r(t)$ to arrive at
\begin{equation*}
S[q] = - \frac{1}{g}   \int \dd t   \left(
\tfrac12   \dot q^2 +
\tfrac12   q^2 - q^4 \right)  , \qquad
s[r] = \frac{1}{g}   \int \dd t   \left(
\tfrac12   \dot r^2 +
\tfrac12   r^2 - r^3 \right)  .
\end{equation*}
Indeed, the width of the resonance is proportional
to the exponential of minus the
Euclidean action of the instanton conf\/iguration,
which in turn is a solution to the classical equations
of motion in the ``inverted'' potentials
$F(q) = q^4 - \tfrac12   q^2$ and
$G(r) = r^3 - \tfrac12   r^2$. The instanton is given in
Fig.~\ref{fig3}. The instanton solutions read
\begin{equation}
\label{inst4}
q(t) = q_{\mathrm{cl}} (t) =
\pm \left[ \cosh(2  t) + 1\right]^{-1/2}
\end{equation}
for the quartic and
\begin{equation}
\label{inst3}
r(t) = r_{\mathrm{cl}}(t) =
\left[ \cosh(t) + 1\right]^{-1}
\end{equation}
for the cubic potential.
Evaluating the instanton action, one obtains the leading-order
results,
\begin{equation*}
S[q_{\mathrm{cl}}] = - \frac{1}{3   g}  , \qquad
s[r_{\mathrm{cl}}] = \frac{2}{15   g}  .
\end{equation*}
Observe that both instanton actions are positive
in the relevant regions, where the potential
is unstable ($g < 0$ and $g > 0$, respectively).
Consequently, the decay widths of the resonances
of the quartic and cubic potential are proportional to
$\exp[1/(3 g)]$ and
$\exp[-2/(15 g)]$, respectively.

In order to evaluate higher-order corrections are general
formulas for oscillators of arbitrary degree,
one needs dispersion relations.
These reads for the quartic and the cubic, respectively,
\begin{subequations}
\begin{equation}
\label{disp4}
E_n(g) = n + \frac12 - \frac{g}{\pi}  \int_{-\infty}^0 \dd s \
\frac{{\rm Im}\, E_n(s+{\rm i}\,0)}{s   (s - g)}  ,
\end{equation}
and~\cite{BeDu1999}
\begin{equation}
\label{disp3}
\epsilon_n(g) = n + \frac12 +
\frac{g}{\pi}  \int_0^{\infty} \dd s
\frac{{\rm Im} \, \epsilon_n^{(M)}(s + {\rm i}\, 0)}{s  (s - g)} .
\end{equation}
\end{subequations}
One might ask if the integration for the cubic
really stretches to $s = +\infty$. The answer is af\/f\/irmative:
according to~\cite{JeSuLuZJ2008}, we may write
the leading terms for the complex strong-coupling expansion for the
f\/irst three resonances of the cubic as
\begin{gather*}
  \epsilon_0(g + {\rm i} \, 0) \;  \mathop{=}^{g \to \infty} \;  g^{1/5}  \,
0.762 851 775 \, \ee^{-\ii \, \pi/5}  ,
\\
  \epsilon_1(g + {\rm i} \, 0) \;  \mathop{=}^{g \to \infty} \;  g^{1/5}\,
2.711 079 923 \, \ee^{-\ii \, \pi/5}  ,
\\
  \epsilon_2(g + {\rm i} \, 0) \;  \mathop{=}^{g \to \infty} \;  g^{1/5} \,
4.989 240 088 \, \ee^{-\ii \, \pi/5}  .
\end{gather*}
Here, we choose boundary
conditions for the wave functions such as to generate
resonance energies with a negative imaginary part,
which are relevant for the dispersion integral~\eqref{disp3}
as they are ``attached'' to values of the
coupling with an inf\/initesimal positive imaginary part.
Intuitively, we might assume that at least the second and
the third resonance might disappear for very strong coupling $g$.
This is because the classically forbidden region
of the cubic potential which separates the fall-of\/f region
from the ``middle valley'' becomes smaller and smaller as the
coupling increases, and indeed, the second excited level lies
well above the relevant energy region in which tunneling would
be necessary (see also Fig.~\ref{fig2}(b)).
However, this point of view does not hold:
the resonance persists for arbitrarily large coupling,
and the physical picture is that the ``escape''
of the probability density to inf\/inity, which for the
cubic happens in f\/inite time, provides for a suf\/f\/icient
mechanism to induce a nonvanishing decay width of the
resonance, even if the traditional tunneling picture is
not applicable (similar considerations apply to the
resonances in the Stark ef\/fect~\cite{BeGrHaSi1979}).
These considerations can be generalized to odd potentials
of arbitrary order, and to arbitrary excited
levels~\cite{JeSuZJ2009prl}. One result stemming from this
generalization is given in the Appendix.

Using generalized Bohr--Sommerfeld quantizations
which are inspired by the treatment of double-well-like
potentials~\cite{ZJJe2004i,ZJJe2004ii},
one can formulate a general formalism~\cite{JeSuZJ2009prl}
which allows to write down higher-order formulas for the
complex resonance energies.
In contrast to~\cite{JeSuZJ2009prl},
where we focused on the f\/irst few correction terms
for the anharmonic oscillators
of the third, sixth and seventh degree,
here we would like to fully concentrate on the
cubic and quartic oscillators and indicate the
generalized nonanalytic expansion exclusively for the
oscillators of the third and the forth degree.
Specif\/ically, we have for a resonance of the quartic,
\begin{subequations}
\label{GENIE}
\begin{gather}
E_n(g < 0) =
\sum_{K=0}^\infty E_{n,K} \, g^K
\nonumber\\
\phantom{E_n(g < 0) = }{}  + \sum_{J=1}^\infty
\left[ \frac{\ii}{\sqrt{\pi} \, n!}
\frac{2^{2   n + \frac12}}%
{(-g)^{n + \frac12}}
\exp\left( -\frac{1}{3   g} \right) \right]^J
\sum_{L=0}^{J-1}
\, \ln^L\left( \frac{4}{g} \right)
\sum_{K=0}^\infty \Xi^{(4,n)}_{J,L,K}   g^K\label{GENIE4}
\end{gather}
and for a general resonance of the cubic,
\begin{gather}
\epsilon_n(g > 0) =
\sum_{K=0}^\infty \epsilon_{n,K} \, g^K
\nonumber\\
\phantom{\epsilon_n(g > 0) =}{}  + \sum_{J=1}^\infty
\left[ \frac{\ii}{\sqrt{\pi} \, n!}
\frac{2^{3   n}}{g^{n + \frac12}}
\exp\left( - \frac{2}{15   g} \right) \right]^J
\sum_{L=0}^{J-1}
  \ln^L\left( - \frac{8}{g} \right)
\sum_{K=0}^\infty \Xi^{(3,n)}_{J,L,K}   g^K   ,\label{GENIE3}
\end{gather}
\end{subequations}
where the $\Xi$ are constant coef\/f\/icients.
(In contrast to~\cite{JeSuZJ2009prl}, we here
single out the perturbative contributions
$\sum_{K=0}^\infty E_{n,K} g^K$ and
$\sum_{K=0}^\infty \epsilon_{n,K}  g^K$ from the
instanton ef\/fects, which are given by the
terms with $J = 1, \ldots, \infty$.)

Of particular phenomenological relevance is the term with
$J = 1$ as it contains the perturbative corrections about the
instanton conf\/iguration and is very important for comparison
with numerically determined
resonance  eigenenergies of the systems.
Without details, we only quote here~\cite{JeSuZJ2009prl} the results for the
higher-order corrections to the ground state and to the
f\/irst excited state of the quartic, which read
\begin{subequations}
\label{higher4}
\begin{gather}
 {\rm Im} \, E_0(g < 0) = - \exp\left( \frac{1}{3\,g} \right)
\sqrt{- \frac{2}{\pi g} }
\left\{ 1
+ \frac{95}{24}   g
- \frac{ 13259 }{ 1152 }   g^2
+ \frac{ 8956043 }{ 82944 }   g^3 \right.
\nonumber\\
\hphantom{{\rm Im} \, E_0(g < 0) =}{}
- \frac{ 11481557783 }{ 7962624 }   g^4
+ \frac{ 4580883830443 }{ 191102976 }   g^5
- \frac{ 12914334973382407 }{ 27518828544 }   g^6
\label{im40}\\
\left.\hphantom{{\rm Im} \, E_0(g < 0) =}{}
+ \frac{ 6938216714164463905 }{ 660451885056 }   g^7
- \frac{ 33483882026182043052421 }{ 126806761930752 }   g^8
+ {\mathcal O}(g^9) \right\}\nonumber
\end{gather}
and
\begin{gather}
 {\rm Im} \, E_1(g < 0) = - \exp\left( \frac{1}{3\,g} \right)
\sqrt{- \frac{32}{\pi g^3} }
\left\{ 1
+ \frac{371}{24}   g
- \frac{ 3371 }{ 1152 }   g^2
+ \frac{ 33467903 }{ 82944 }   g^3
\right.
\nonumber\\
\phantom{{\rm Im} \, E_1(g < 0) =}{}
- \frac{ 73699079735 }{ 7962624 }     g^4
+ \frac{ 44874270156367 }{ 191102976 }    g^5
- \frac{ 181465701024056263 }{ 27518828544 }    g^6
\label{im41}\\
\left.\phantom{{\rm Im} \, E_1(g < 0) =}{}
+ \frac{ 133606590325852428349 }{ 660451885056 }   g^7
- \frac{ 850916613482026035123397 }{ 126806761930752 }    g^8
+ {\mathcal O}(g^9) \right\}  ,\nonumber
\end{gather}
\end{subequations}
and for the lowest two levels of the cubic, which are
\begin{subequations}
\label{higher3}
\begin{gather}
 {\rm Im} \, \epsilon_0(g > 0) =
-\frac{\exp\left( - \frac{2}{15\,g} \right)}%
{\sqrt{\pi   g}}
\left\{ 1 - \frac{169}{16}   g -
\frac{ 44507 }{ 512 }   g^2 -
\frac{ 86071851 }{ 40960 }   g^3 -
\frac{ 189244716209 }{ 2621440 }   g^4  \right.
\nonumber\\
 \phantom{{\rm Im} \, \epsilon_0(g > 0) =}{}  - \frac{ 128830328039451 }{ 41943040 }   g^5  -
\frac{ 1027625748709963623 }{ 6710886400 }   g^6
\label{im30}\\
\left.\phantom{{\rm Im} \, \epsilon_0(g > 0) =}{}
- \frac{ 933142404651555165943 }{ 107374182400 }   g^7  -
\frac{ 7583898146256325425743381 }{ 13743895347200 }   g^8
+ {\mathcal O}(g^9) \right\}\nonumber
\end{gather}
and
\begin{gather}
{\rm Im} \, \epsilon_1(g > 0) =
- \frac{ 8   \exp\left( - \frac{2}{15 g} \right) }%
{ \sqrt{\pi}   g^{3/2}}
\left\{
1 - \frac{853}{16}   g
+ \frac{33349}{512}   g^2 - \frac{395368511}{40960}   g^3
\right.
\nonumber\\
\phantom{{\rm Im} \, \epsilon_1(g > 0) =}{}  - \frac{ 1788829864593 }{ 2621440 }   g^4
- \frac{ 2121533029723423 }{ 41943040 }   g^5\nonumber\\
 \phantom{{\rm Im} \, \epsilon_1(g > 0) =}{}- \frac{ 27231734458812207783 }{ 6710886400 }   g^6
- \frac{ 37583589061337851179291 }{ 107374182400 }   g^7\nonumber\\
\left.\phantom{{\rm Im} \, \epsilon_1(g > 0) =}{}
- \frac{ 442771791224240926548268373 }{ 13743895347200 }   g^8
+ {\mathcal O}(g^{9}) \right\}  .\label{im31}
\end{gather}
\end{subequations}
Note that the higher-order terms for the ground state of the
cubic, by virtue of the dispersion relation (\ref{disp3}),
are in full agreement with the higher-order formulas
given in~\cite{BeDu1999}.
Note also that both above results for the ground state could have been
found by plain perturbation theory about the instanton
conf\/iguration, but the results for the
excited states are somewhat less trivial to obtain;
they follow from the general formalism outlined in~\cite{JeSuZJ2009prl}.

\section{Conclusions}
\label{conclu}

Our generalized nonanalytic expansions (\ref{GENIE4}) and
(\ref{GENIE3}) provide for an accurate description of resonance
energies of the quartic and cubic anharmonic oscillators.
These combine exponential factors, logarithms and power series
in a systematic, but highly nonanalytic formula.
Note that the term ``resurgent functions'' has been used in
the mathematical literature~\cite{Ph1989,DeDiPh1990,CaNoPh1993}
in order to describe such mathematical structures;
we here attempt to denote them using a more descriptive,
alternative name.

In a general context, we conclude that
a physical phenomenon sometimes cannot be described
by a power series alone. We have to combine more than one functional
form in order to write down a systematic, but not necessarily analytic
expansion in order to describe the phenomenon in question.
In the context of odd anharmonic oscillators,
the generalized nonanalytic expansions which describe the
energy levels in higher orders are intimately connected to the
dispersion relations (\ref{disp4}) and (\ref{disp3})
which in turn prof\/it from the $\mathcal{PT}$-symmetry
of the odd anharmonic oscillators for purely imaginary coupling.
The $\mathcal{PT}$-symmetry is used here as an indispensable,
auxiliary device in our analysis (it is perhaps interesting to note that
the use of $\mathcal{PT}$-symmetry as an auxiliary device has recently
helpful in a completely dif\/ferent context~\cite{AnCaGiKaRe2008}).
In our case, very large coef\/f\/icients are obtained for, e.g.,
the perturbation about the instanton for the f\/irst
excited state of the cubic (see equation~(\ref{im31})).
At a coupling of $g = 0.01$, the
f\/irst correction term $-853g/16$ halves the result for the
decay width of the f\/irst excited state, and the higher-order
terms are equally important.

We have recently
generalized the above
treatment to higher-order corrections to anharmonic oscillators
up to the tenth order. The oscillators of degree six and seven
display very peculiar properties: for the sixth degree, some of
the correction terms accidentally cancel, and for the
septic potential, the corrections can be expressed in a natural
way in terms of the golden ratio $\phi = (\sqrt{5} + 1)/2$.
For the potential of the seventh degree,
details are discussed in~\cite{JeSuZJ2009prl}.

Let us conclude this article with two remarks
regarding the necessity of using general non-analytic
expansions to describe physical phenomena.
First, the occurrence of the nonanalytic exponential
terms is connected with the presence of branch cuts
relevant to the description of physical quantities
as a function of the coupling, as exemplif\/ied by the
equations~\eqref{disp4} and~\eqref{disp3}. Second, the
presence of higher-order terms in the generalized expansions
is due to
our inability to solve the eigenvalue equations exactly, or,
in other words, to carry out
WKB expansions in closed form to arbitrarily high order.
These two facts, intertwined, give rise to the
mathematical structures that we f\/ind here in
equations~\eqref{GENIE4} and~\eqref{GENIE3}.

\pdfbookmark[1]{Appendix}{app}
\section*{Appendix}

Using the dispersion relation (\ref{disp3})
and a generalization of the instanton conf\/iguration
(\ref{inst3}) to arbitrary odd oscillators,
one may evaluate the decay width for a general state
of an odd potential and general
large-order (``Bender--Wu'') formulas
for the large-order behavior of the
perturbative coef\/f\/icients of arbitrary excited levels
for odd anharmonic oscillators.
For a general perturbation of the form $\sqrt{g}\,x^M$,
with odd $M \geq 3$, with resonance energies
$\epsilon^{(M)}_n(g) \sim \sum_K \epsilon^{(M)}_{n,K}   g^K$,
we obtain~\cite{JeSuZJ2009prl} in the limit $K \to \infty$,
\begin{gather*}
\epsilon^{(M)}_{n,K} \;  \sim  \;
-\frac{( M - 2 )   \Gamma\left( (M - 2) K + n + \tfrac12 \right)}%
{\pi^{3/2}   n!   2^{2 K + 1 - n}}
\left[ B\left(\frac{M}{M-2}, \frac{M}{M-2}\right)
  \right]^{-(M - 2) K - n - \tfrac12}  ,
\end{gather*}
where $B(x,y) = \Gamma(x)   \Gamma(y) / \Gamma( x + y )$
is the Euler Beta function.

\subsection*{Acknowledgments}

U.D.J.~acknowledges helpful conversations with
C.M.~Bender and J.~Feinberg at PHHQP2008 at the
conference venue in Benasque (Spain).
A.S.~acknowledges support from the
Helmholtz Gemeinschaft (Nachwuchsgruppe VH--NG--421).

\pdfbookmark[1]{References}{ref}

\LastPageEnding

\end{document}